# Anisotropic linear and nonlinear charge-spin conversion in topological semimetal SrIrO$_3$

Bin Lao[1,2,7], Peitao Liu[3,7], Xuan Zheng[1,2,4,7], Zengxing Lu[1,2], Sheng Li[1,2], Kenan Zhao[1,2], Liguang Gong[1,2], Tao Tang[1,2], Keyi Wu[1,2], You-guo Shi[5], Yan Sun[3], Xing-Qiu Chen[3], Run-Wei Li[1,2,6] and Zhiming Wang[1,2,6*]

[1]CAS Key Laboratory of Magnetic Materials and Devices, Ningbo Institute of Materials Technology and Engineering, Chinese Academy of Sciences, Ningbo 315201, China
[2]Zhejiang Province Key Laboratory of Magnetic Materials and Application Technology, Ningbo Institute of Materials Technology and Engineering, Chinese Academy of Sciences, Ningbo 315201, China
[3]Shenyang National Laboratory for Materials Science, Institute of Metal Research, Chinese Academy of Sciences, Shenyang 110016, China
[4]New Materials Institute, University of Nottingham Ningbo China, Ningbo 315100, China
[5]Institute of Physics, Chinese Academy of Sciences, Beijing 100190, China
[6]Center of Materials Science and Optoelectronics Engineering, University of Chinese Academy of Sciences, Beijing 100049, China
[7]These authors contributed equally: Bin Lao, Peitao Liu, Xuan Zheng.
*Email: zhiming.wang@nimte.ac.cn

## Abstract

Over the past decade, utilizing spin currents in the linear response of electric field to manipulate magnetization states via spin-orbit torques (SOTs) is one of the core concepts for realizing a multitude of spintronic devices. Besides the linear regime, recently, nonlinear charge-spin conversion under the square of electric field has been recognized in a wide variety of materials with nontrivial spin textures, opening an emerging field of nonlinear spintronics. Here, we report the investigation of both linear and nonlinear charge-spin conversion in one single topological semimetal SrIrO$_3$(110) thin film that hosts strong spin-orbit coupling and nontrivial spin textures in the momentum space. In the nonlinear regime, the observation of crystalline direction dependent response indicates the presence of anisotropic surface states induced spin-momentum locking near the Fermi level. Such anisotropic spin textures also give rise to spin currents in the linear response regime, which mainly contributes to the fieldlike SOT component. Our work demonstrates the power of combination of linear and nonlinear approaches in understanding and utilizing charge-spin conversion in topological materials.

# Introduction

The generation and manipulation of spin currents is one of the crucial aspects in the field of spintronics. Spin currents converted by charge currents exert spin-orbit torques (SOTs) on the adjacent magnetic material through the transfer of angular momentum, leading to current-induced magnetization switching with low energy consumption in spintronic devices [1-2]. In general, charge-spin conversion can in principle arise from spin-orbit coupling (SOC), including bulk spin Hall effect (SHE), and nontrivial spin textures at the surface/interface due to inversion symmetry broken, such as spin-momentum locking [3-9]. In the presence of these mechanisms, the efficiencies of charge-spin conversion (referred as SOT-efficiencies) are widely accepted and reasonably evaluated in the linear response regime [4, 10-14]. Under a small perturbation, such as an external electric field, both distribution and eigenstate in a carrier system are modified, giving rise to different nonequilibrium distributions and thereby the spin currents. Besides the linear response regime, recently, high-order response of charge-spin conversion to electric field has been theoretically proposed and also experimentally observed in noncentrosysmetric transition metal dichalcogenides, topological Dirac semimetals, and two-dimensional Rashba-Dresselhaus systems [15-30]. The high-order response is closely related to nontrivial spin textures and topology due to strong SOC with inversion symmetry broken, resulting in a variety of quantum phenomena, such as nonlinear planar/anomalous Hall, nonreciprocal nonlinear optical and inverse Edelstein effects. Conversely, these exotic phenomena under the nonlinear responses also becomes a powerful probe to capture the nontrivial spin textures properties, such as spin-momentum locked surface states via magnetoelectric transport studies [26-28]. Thus the ubiquitous nonlinear response effects in materials systems with inversion asymmetry hold tremendous potential for adding new functionalities and characterization means in the current spintronics.

Transition metal oxides exhibit remarkable and highly tunable magnetoelectronic as well as SOC induced properties, due to the strong coupling between charge, spin, orbital and lattice degrees of freedom, offering a versatile platform to pursue desirable functionality for next-generation spintronic devices [31-46]. In particular, 5d transition-metal oxide $SrIrO_3$ exhibits pronounced charge-spin conversion in both linear and nonlinear response [47-52]. In the linear regime, large SOT-efficiencies has been suggested to be associated with the bulk SHE, stemming from strong SOC related topological nature in $SrIrO_3$ [14, 41-46]. However, the dependence of SOT-efficiencies on crystalline direction/orientation shows discrepancies among the previous studies, implicating the underlying mechanisms in the linear charge-spin conversion. In the nonlinear regime, an observation of the nonlinear planar Hall effect in $SrIrO_3$(001) evidences the existence of novel surface states with spin-momentum locking, which was ascribed to the strain induced nontrivial spin textures [28]. In addition, theoretical calculations predict that such surface states should commonly exist in perovskite $SrIrO_3$ with varied crystal orientations protected by lattice symmetry [47-48]. Moreover, since the spin-momentum locking was reported to contribute large SOT-efficiencies in topological insulators and transition metal dichalcogenides [6-9], it may also play a role in the linear response in $SrIrO_3$. The outstanding charge-spin conversion features and the entanglement of the associated microscopic mechanisms therefore call for a comprehensive investigation to get deeper insight into the order-relevant charge-spin conversion.

In this work, we experimentally investigate the linear and nonlinear charge-spin conversion in SrIrO$_3$(110) thin films. The perovskite SrIrO$_3$ has an orthorhombic crystal structure with *Pbnm* space group. The lattice constants are $a$ = 5.60 Å, $b$ = 5.58 Å, and $c$ = 7.89 Å, corresponding to a pseudocubic structure of lattice constant $a_{pc}$ = 3.95 Å, so the SrIrO$_3$ thin films can be epitaxially grown on several substrates with small lattice mismatch, such as SrTiO$_3$ [53-55]. For SrIrO$_3$ growing on SrTiO$_3$(110), figure 1(a) shows top view of the crystal structure along the [110] direction, where the Sr atoms and IrO$_6$ octahedrons align anisotropically in the *xy* plane, resulting in two types of mirror symmetry with respect to in-plane [001] and [1-10] directions. Since the predicted spin-momentum locking and evaluated SOTs are closely related to the crystalline directions [28, 41-42, 47-48], the highly anisotropic properties of SrIrO$_3$(110) for current along [1-10] and [001] directions enable us to explore both response behaviors simultaneously. Specifically, as shown in figure 1(b) and (c), we used the nonlinear planar Hall measurement to detect the spin-momentum locking induced second-order transverse charge-spin current in epitaxial thin film SrIrO$_3$(110); And in heterostructures of SrIrO$_3$/NiFe, harmonics Hall measurements were carried out to characterize the spin currents induced SOTs in the linear response regime. As a result, both coefficients of nonlinear planar Hall effect and SOT-efficiencies exhibit strong crystalline direction dependence. The coefficients can be twofold different along the [1-10] and [001] directions. These strongly anisotropic responses are understood from first-principles calculations of surface states and spin Hall conductivities. Combining the experimental results and theoretical calculations, we conclude that the anisotropic nonlinear charge-spin conversion associated with the strength of spin-momentum locking arise from the specific spin textures at the Fermi surface. Furthermore, such spin-momentum locking also acts as a source in the case of the linear response regime, which dominates the interfacial contribution of SOT-efficiencies. Our study links the linear and nonlinear charge-spin conversion in one single topological semimetal SrIrO$_3$, disentangling the mechanisms of its prominent conversion efficiencies. Since the order-relevant relationship should be general in more diverse materials with strong SOC and nontrivial spin textures, our findings thus provide a powerful method in interpreting and exploiting emergent charge-spin conversion phenomena for linear and nonlinear spintronics.

## Results

**Nonlinear planar Hall measurements**

Figure 2(a) shows the schematic diagram of the nonlinear planar Hall measurement, in which angular dependence of second harmonic transverse Hall voltage $V_{xy}^{2\omega}$ and longitudinal voltage $V_{xx}^{2\omega}$ on the in-plane magnetic field direction with respect to the current direction in the SrIrO$_3$ single layer were measured. For a fixed AC current *I* along the [001] direction of 0.8 mA and a magnetic field *H* of 4 T at temperature of 290 K, as an example shown in Fig. 2(b), a typical second harmonic planar Hall resistance $R_{xy}^{2\omega} = V_{xy}^{2\omega}/I$ as a function of the angle $\varphi$ exhibits cosine dependence. This suggests nonlinear planar Hall effect (NHE) associated with complex spin textures, as reported in the SrIrO$_3$(100) films [28]. Note that the cos$\varphi$ angular dependence can be ascribed to other contributions. For instance, firstly, Nernst effect induced by a thermal gradient can produce such angular dependence and yield isotropic Nernst resistivity between transverse and longitudinal directions ($\rho_{xy}^{2\omega}/\rho_{xx}^{2\omega} = 1$) in the in-plane magnetic field scan. We estimated the ratio $\rho_{xy}^{2\omega}/\rho_{xx}^{2\omega} = (R_{xy}^{2\omega}/R_{xx}^{2\omega})(L/W)$ to be about 1.43, indicating that the Nernst effect is NOT the dominant contribution in our measurements (see Supplemental Material S1). Secondary, asymmetric magnon-mediated scattering can also generate the $\cos\varphi$ dependence, as reported in a magnetic-nonmagnetic topological system [25]. However, this contribution can be readily ruled out in our SrIrO$_3$ sample due to the absence of electron-magnon scattering from a magnetic layer.

To uncover the observed nonlinear nature, angular dependent second harmonic planar Hall signals under various magnitudes of current and magnetic field were further performed. After subtracting a constant background component of the signals, the second harmonic planar Hall resistance amplitudes $\Delta R_{xy}^{2\omega}$ as a function of $\varphi$ for a series magnitudes of $I$ under a fixed $H$ of 4 T are obtained and plotted in Fig. 2(c). Qualitatively, all the curves exhibit $\cos\varphi$ shape and $\Delta R_{xy}^{2\omega}$ increases monotonously with increasing $I$ from 0.2 to 1.5 mA. Figure 2(d) exhibits the same tendency of $\Delta R_{xy}^{2\omega}$ at the $H$ of 0.5 to 4 T with a fixed $I$ of 0.8 mA. We further confirmed that the extracted $\Delta R_{xy}^{2\omega}$ exhibits bilinear dependence on the magnitudes of the current $I$ and magnetic field $H$, as shown in Fig. 2(e) and 2(f), unambiguously evidencing the nonlinear nature of $R_{xy}^{2\omega}$. This bilinear behavior is a distinctive feature of the nonlinear planar Hall effect, which reflects a spin-momentum locking phenomenon introduced by the complex spin textures in momentum space. This phenomenon can be understood by the fact that in the presence of an electrical field $E$ due to the applied $I$, the spin-momentum locked surface states generate spin fluxes of nonlinear spin current with opposite momentum at the second-order of $E$. By applying the in-plane $H$ simultaneously, the balance of the spin fluxes is broken, resulting in a partial conversion of the spin current into a charge current with direction orthogonal to the $H(\varphi)$ [26-28]. Thus, the transverse $R_{xy}^{2\omega}\cos\varphi$ and longitudinal $R_{xx}^{2\omega}\sin\varphi$ second harmonic signals arise, and the larger the $I$ and $H$, the stronger the amplitude of the signals. To quantitatively estimate the strength of spin-momentum locking, we computed the coefficient of the nonlinear planar Hall resistivity generated by per unit electric and magnetic field through $\Delta\rho_{xy}^{2\omega}/(E_xH)$, where $E_x = I R_{xx}^{1\omega}/L$ is the longitudinal electrical field collinear with the $I$. Substitution of $\Delta\rho_{xy}^{2\omega} = 3.3 \times 10^{-3}$ mΩ μm, $V_{xx}^{1\omega}= I R_{xx}^{1\omega} = 694$ mV with $I = 0.8$ mA and $H = 4$ T, the $\Delta\rho_{xy}^{2\omega}/(E_xH)$ value is calculated to be 0.060 mΩ μm$^2$ ·V$^{-1}$ ·T$^{-1}$ at temperature of 290 K, which is comparable to the most recently reported values of SrIrO$_3$ deposited on other substrates at room temperature [28].

**Spin-orbit torques measurements**

Spin-orbit torques (SOTs) arise from angular momentum transfer by spin currents via bulk and interfacial contributions, and have two components of dampinglike torque $\tau_{DL}$ ($\approx M \times H_{DL}$) and fieldlike torque $\tau_{DL}$ ($\approx M \times H_{FL}$), where the $H_{DL}$ and $H_{FL}$ are the equivalent magnetic fields generated by the spin currents. The SOTs can exert on magnetization $M$ of an adjacent magnet and then force the $M$ to precess [56-58]. Hence, detecting such precession enables us to quantitatively study the DL- and FL-SOT efficiencies. The SOT-efficiencies of SrIrO$_3$(110) were evaluated in the SrIrO$_3$/NiFe patterned Hall-bar device by detecting the magnetization precession of the NiFe through the spin currents produced from the SrIrO$_3$ using harmonic Hall measurements [9, 59-60]. Figure 3(a) schematically presents the direction of the in-plane magnetic field $H$, applied AC current $I$ and acquired voltage signals in our measurement configuration. All measurements were performed at 290 K. The first-harmonic Hall voltage $V_{xy}^{1\omega}$ (the so-called planar Hall voltage) measured on the SrIrO$_3$/NiFe device follows the sin2$\varphi$ dependence, as shown in Fig. 3(b), suggesting that the magnetization of the NiFe is always aligned with the in-plane magnetic field for $H > 300$ Oe. The second harmonic Hall voltage $V_{xy}^{2\omega}$ was measured simultaneously with the $V_{xy}^{1\omega}$. The $V_{xy}^{2\omega}$ contains both $H_{DL}$ and $H_{FL}$ components, which can be analyzed by the following relationship when the magnetization of the NiFe sufficiently lies in plane:

$$V_{xy}^{2\omega} = \tfrac{1}{2}V_{DL}\cos\varphi + (-V_{FL}\cos\varphi\cos2\varphi), \tag{1}$$

$$V_{DL} = \frac{V_{AHE}}{H_K - H}H_{DL}, \tag{2}$$

$$V_{FL} = \frac{V_{PHE}}{H}(H_{FL} + H_{Oersted}). \tag{3}$$

Here, $V_{DL}$ is the dampinglike voltage containing the $H_{DL}$ which follows the same cos$\varphi$ dependence, and $V_{HL}$ is the fieldlike voltage including the contributions of the $H_{FL}$ and an Oersted field $H_{Oersted}$ in the SrIrO$_3$ layer with cos$\varphi$cos2$\varphi$ angular dependence; $H_K = 0.5$ T is the effective anisotropy field of the NiFe layer obtained from an anomalous Hall effect measurement, $V_{AHE}$ and $V_{PHE}$ ($= V_{xy}^{1\omega}$) are the anomalous Hall voltage and the planar Hall voltage, respectively. By fitting a representative $V_{xy}^{2\omega}$ data with $I = 2.5$ mA and $H = 300$ Oe as shown in Fig. 3(c) to Eq. (1), the experimental data were well reproduced by considering both dampinglike and fieldlike contributions. Figures 3(d) and 3(e) display the obtained $V_{DL}$ and $V_{HL}$ for $I$ varying from 1.5 to 4.5 mA as a function of $1/(H + H_K)$ and $1/H$, respectively. All the $V_{DL}$ and $V_{HL}$ data for each $I$ follow a linear relationship against the abscissa. Accordingly, the slopes of these linear fitting curves correspond to the $H_{DL}V_{AHE}$ and the $(H_{FL}+H_{Oersted})V_{PHE}$, respectively. To estimate $H_{FL}$, $H_{Oersted} = \mu_0 J t_{SIO}/2$ was excluded from the $H_{FL}+H_{Oersted}$ term based on Ampere's law, where $J$ denotes the current density flowing into the SrIrO$_3$ layer calculated based on current-shunting effect and $t_{SIO} = 25$ nm is the thickness of the SrIrO$_3$. Similarly, by substituting the value of $V_{AHE}$, $H_{DL}$ were estimated quantitatively and plotted together with the $H_{FL}$ against $J$ shown in Fig. 3(f). The effective fields per current density $H_{DL}/J$ and $H_{FL}/J$ were determined to be $-35.3\pm1.1$ Oe/($10^{11}$ A/m$^2$) and $-22.2\pm0.1$ Oe/($10^{11}$ A/m$^2$). As the effective fields are attributed to the dampinglike and fieldlike torques, we finally estimated the DL- and FL-SOT efficiencies by using the formula $\eta_{DL(FL)} = (2e\mu_0 M_s t_{NiFe}/\hbar)(H_{DL(FL)}/J)$ [42, 61], where $M_s = 610$ emu/cm$^3$ and $t_{NiFe} = 7$ nm are the saturation magnetization and the thickness of the NiFe layer, respectively (see Methods). The calculated absolute value of $\eta_{DL}$ is about 0.46 and $\eta_{FL}$ is about 0.29. Note that the $\eta_{DL}$ is comparable to the reported values of SrIrO$_3$ [41-45], especially closed to the results of a SrIrO$_3$/NiFe system [41, 45].

**Anisotropic NHE and SOTs**

To further explore the linear and nonlinear charge-spin conversion of spin-momentum locking in SrIrO$_3$ (110), we evaluated and compared the NHE and SOTs with current along

[001] and [1-10] directions, respectively. Since the complex spin textures near the Fermi surface is dependent to spatial symmetry of lattice, both the NHE and SOTs induced by the spin momentum locked surface states should be highly anisotropic and significantly different for each crystalline direction. In the above sections, the results of the NHE coefficient, as well as the SOT-efficiencies were introduced when an AC current $I$ was applied along the [001] direction. Accordingly, similar measurements for the current flowing along the [1-10] direction were performed. Figure S2 shows the detailed NHE results under various $H$ for $I$//[1-10]. Subsequently, the obtained NHE amplitudes $\Delta R_{xy}^{2\omega}$ at 290 K were plotted together in Fig. 4(a), which increase linearly with $H$ for $I$ flowing along both [001] and [1-10] directions. By a linear fitting, we find the slope of $I$//[001] is about twofold as large as that of $I$//[1-10], implying that the surface states have strongly anisotropic spin textures. It should be stressed here that since the acquired NHE signals reflect the nonlinear transverse spin currents orthogonal to the applied charge currents in a planar Hall geometry, the larger slope of $I$//[001] than for $I$//[1-10] indicates that the spin-momentum locking along the $SrIrO_3$[1-10] is stronger than that along the $SrIrO_3$[001]. By substituting the relevant parameters, the NHE coefficient $\Delta\rho_{xy}^{2\omega}/(E_xH)$ is estimated to be 0.024 mΩ·μm$^2$·V$^{-1}$·T$^{-1}$ for $SrIrO_3$[001]. On the other hand, the linear charge-spin conversion induced effective fields $H_{DL}$ and $H_{FL}$ for $I$//[1-10] were measured by the harmonic Hall method and estimated using Eqs.(1)-(3) at 290 K (see Supplemental Material S3). The results of the effective fields for $I$//[1-10] and $I$//[001] were plotted together as a function of current density $J$, as shown in Fig. 4(b). Consistent with the anisotropy revealed in the NHE measurement, the $H_{DL}$ and $H_{FL}$ exhibit linear dependence with $J$ in both crystalline directions while display anisotropic magnitudes. The absolute values of SOT efficiencies $\eta_{DL}$ and $\eta_{FL}$ for $SrIrO_3$[1-10] are about 0.25 and 0.36, respectively. To further summarize the NHE coefficient and SOT-efficiencies results as shown in Fig. 4(c), we find the FL-SOT efficiency and the strength of the spin-momentum locking for $I$//[1-10] is stronger than that for $I$//[001], whereas the DL-SOT efficiency for $I$//[001] is more remarkable.

## Discussion

Since the SOT-efficiencies exhibit strong anisotropy for the charge current $I$ along the [001] and [1-10] directions in $SrIrO_3$(110), their associated bulk SHE and interfacial spin-momentum locked surface states should also be anisotropic. From the NHE measurement, we confirmed the anisotropic magnitude of $\Delta\rho_{xy}^{2\omega}/(E_xH)$, which is about twofold larger for $I$//[1-10] than for $I$//[001], indicating a stronger spin-momentum locking for $I$//[1-10], consistent with the anisotropy of FL-SOT. As the FL-SOT is mostly contributed by the interfacial spin accumulation, the same anisotropy between the NHE results and FL-SOT suggests that the spin-momentum locking is a source of the FL-SOT. In contrast, DL-SOT exhibits an opposite anisotropy as compared to the FL-SOT. Since the bulk SHE governs the DL-SOT, one would thus expect that the opposite tendency of DL-SOT and FL-SOT may result from the dominant contribution of an anisotropic spin Hall conductivity.

To uncover the underlying mechanisms for the experimentally derived anisotropic behavior in linear and nonlinear response regimes, we performed first-principles calculations of surface states and spin Hall conductivities. Figure 5 shows the compiled results, where a unit cell of SrIrO$_3$(110) and the Brillouin zones are depicted in Figs. 5(a) and 5(b), respectively. It has been reported that the perovskite oxide SrIrO$_3$ exhibit a nonmagnetic correlated state [51, 62] combined with a topological crystalline semimetal character [47-48] and large intrinsic SHE [14, 41, 46]. In line with literature results, the calculated band structure of the (110) surface shows typical surface states close to the Fermi energy (Fig. 5c), which are protected by the crystalline mirror symmetry [51, 62]. By inspecting the Fermi surface, one can readily observe the strongly anisotropic spin-momentum locked surface states between $\bar{\Gamma}$-$\bar{X}$ and $\bar{\Gamma}$-$\bar{Z}$ directions (see Fig. 5d). Notably, the spin-momentum locking is significant around $\bar{X}$ along the $\bar{\Gamma}$-$\bar{X}$ direction (the [1-10] direction in real space), consistent with the nonlinear Hall measurement and anisotropy of FL-SOT. In addition, the calculated spin Hall conductivity (SHC) hosts significant anisotropy with a larger SHC for the charge current along [001] than along [1-10], as shown in Fig. 5(e). This strong anisotropy of SHC may explain the anisotropy of the DL-SOT, i.e. the larger and opposite contribution from the anisotropy of SHC over the anisotropy of surface states lead to the observed anisotropic DL-SOT. Therefore, despite hosting large anisotropic spin-momentum locked surface states evidenced by nonlinear Hall measurement, they mainly contribute to the FL-SOT, while the DL-SOT has the larger contribution from the bulk SHE.

To conclude, we have investigated the linear and nonlinear charge-spin conversion in SrIrO$_3$(110) from combined nonlinear planar Hall measurements, spin-orbit torque measurements and first-principles calculations. In the nonlinear response regime, crystalline direction dependent spin-momentum locking is observed via evaluating the nonlinear planar Hall effect, which reflects an anisotropic spin texture in the surface states. While in the linear response regime, the DL-SOT and FL-SOT exhibit opposite anisotropy of crystalline directions. Comparing the experimental observations and theoretical calculations, we distinguish that the spin-momentum locked surface states dominates the FL-SOT, while the bulk SHE mainly contributes to the DL-SOT. Our work represents a step forward for disentangling the mechanism of high charge-spin conversion efficiencies in topological materials, demonstrating the power of combination of linear and nonlinear approaches in understanding and utilizing charge-spin conversion.

## Methods

**Film and device fabrication**

The SrIrO$_3$(25 nm) single layer and SrIrO$_3$(25 nm)/NiFe(7 nm) bilayer samples were grown on the (110)-oriented SrTiO$_3$ substrates by pulsed laser deposition using a KrF excimer laser ($\lambda$ = 248 nm). During the growth of SrIrO$_3$ the substrate temperature was kept at 700 ºC, and the oxygen partial pressure was 0.1 mbar. After SrIrO$_3$ deposition, the samples were cooled with 20 ºC/min at 1 mbar oxygen pressure. Subsequently, the NiFe layer was deposited on SrIrO$_3$ at room temperature in vacuum for the bilayer sample.

For electric transport measurements, the samples were fabricated into Hall bar patterns by standard photolithography and Ar ion etching techniques. Here, the current path with dimension of 10 μm in width $W$ aligned to SrIrO$_3$[001] and SrIrO$_3$[1-10] directions, respectively. The the length $L$ between two adjacent voltage paths is 50 μm.

**Measurements**

The thickness of the SrIrO$_3$ layer $t_{SIO}$ was determined by an *in-situ* RHEED monitoring during growth. The saturation magnetization $M_s$ and the thickness $t_{NiFe}$ of the NiFe layer were confirmed by a SQUID and a XRR scan, respectively. The electric transport properties were measured using a home-build low-temperature and high magnetic field magneto-electric system. The input AC currents generated by a Keithley6221 with a frequency of 133.73 Hz and varied amplitudes. The output voltages, including the first/second planar Hall voltage $V_{PHE}$ (= $V_{xy}^{1\omega}$) and $V_{xy}^{2\omega}$, first/second magnetoresistance voltage $V_{xx}^{1\omega}$ and $V_{xx}^{2\omega}$, and anomalous Hall voltage $V_{AHE}$, were acquired by lock-in SR 830 (Stanford). The anisotropy field $H_k$ of NiFe was determined by the $V_{AHE}$ measurement. The angular dependence of the magnetotransport voltages were measured with a sample rotator.

**First-principles calculations**

First-principles calculations were performed using the Vienna *ab initio* Simulation Package (VASP) [63-64]. The revised Perdew-Burke-Ernzerhof (PBEsol) functional [65] was used, since it yields a better description of densely-packed structures than the PBE functional [66]. The spin-orbit coupling (SOC) was included. A Hubbard interaction of U = 1.39 eV calculated by the constrained random phase approximation [67] was imposed on Ir-5d orbitals within the DFT+U framework [68]. For all calculations, the plane-wave cutoff for the orbitals was set to 520 eV and the Brillouin zone was sampled with a *k*-point density of 0.13 Å$^{-1}$. The crystal structure was fully relaxed until the Hellmann-Feynman forces acting on each atom were less than 5 meV/Å. The surface electronic structures were computed using the WannierTools code [69]. The Wannier functions and spin Hall conductivities were obtained using the Wannier90 suite [70-71].

## Data Availability

Data are available from the corresponding author upon reasonable request.

## Acknowledgments

This work was supported by the National Key Research and Development Program of China (Nos. 2017YFA0303600, 2019YFA0307800), the National Natural Science Foundation of China (Nos. 12174406, 11874367, 51931011, 52127803), the Key Research Program of Frontier Sciences, Chinese Academy of Sciences (No. ZDBS-LY-SLH008), K.C.Wong Education Foundation (GJTD-2020-11), the 3315 Program of Ningbo, the Natural Science Foundation of Zhejiang province of China (No. LR20A040001), the Beijing National Laboratory for Condensed Matter Physics.

## Author contributions

Z.W. supervised the research; B.L. performed sample growth, device fabrication, transport measurements and data analysis; X.Z., Z.L., S.L., L.G. and K.Z. helped sample growth and device fabrication; P.L. performed first-principles calculations and data analysis with the help of Y. S and X. C.. B.L., P.L. and Z.W. wrote the manuscript. All authors participated in discussions and commented on the manuscript.

## Competing interests

The authors declare no competing interests.

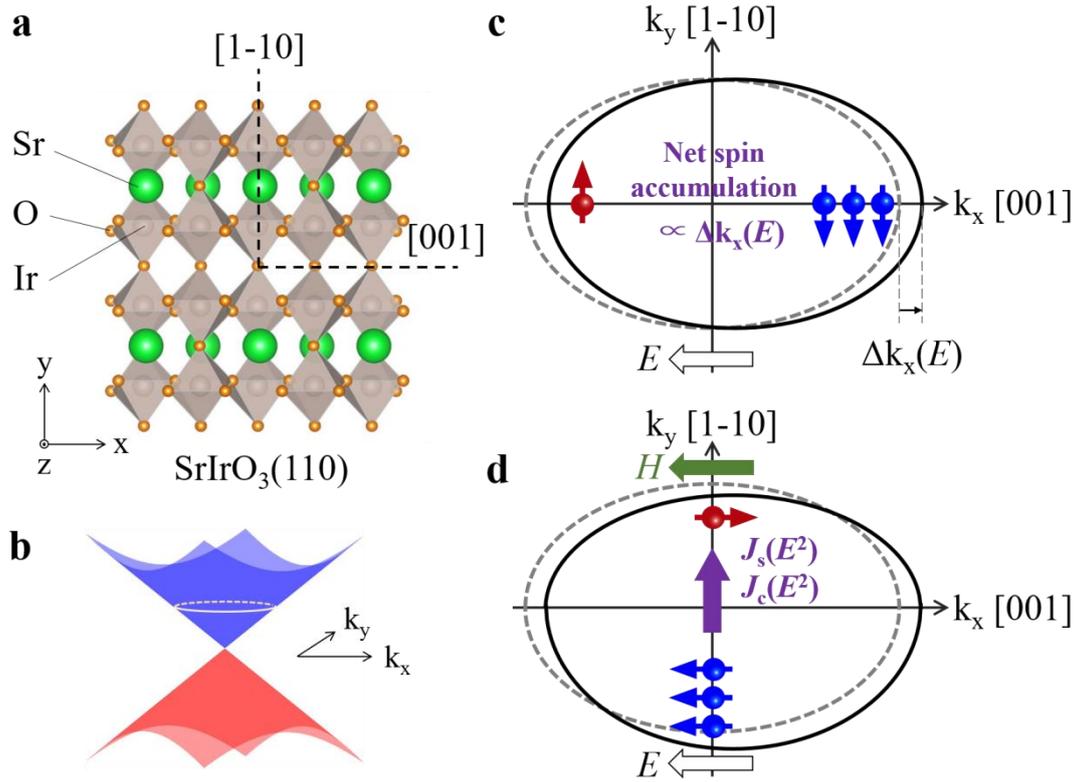

**Fig. 1 Schematic diagrams of the anisotropic linear and nonlinear charge-spin conversion in SrIrO$_3$(110). a** Top view of the SrIrO$_3$(110) crystalline structure along the [110] direction. The dash lines are the two orthogonal crystalline directions of [001] and [1-10]. **b** Schematic view of band structure in the k$_x$-k$_y$ plane near the Dirac node. The circle denotes the anisotropic Fermi surface at the Fermi level. **c** The linear charge-spin conversion in the anisotropic Fermi surface, by which an electric field $E$ (equivalent to a charge current) applied along the [001] direction induces a shift of the Fermi surface, resulting in a net spin accumulation with respect to the shift $\Delta k_x(E)$. Then the spin accumulation couples to the adjacent magnetic layer, leading to a magnetization precession due to SOTs, and can be detected via the harmonics Hall measurements. **d** Illustration of change-spin conversion in nonlinear regime. Based on the configuration of Fig. 1(c), a nonlinear spin current $J_s(E^2)$ along the [1-10] direction is generated simultaneously. When a magnetic field $H$ applied along the [001] direction (parallel to $E$), a further asymmetric distortion of the Fermi surface occurs along the [1-10] direction, and the $J_s(E^2)$ is partially converted into a charge current $J_c(E^2)$ perpendicular to $E$. The $J_c(E^2)$ can be detected via the nonlinear planar Hall measurement.

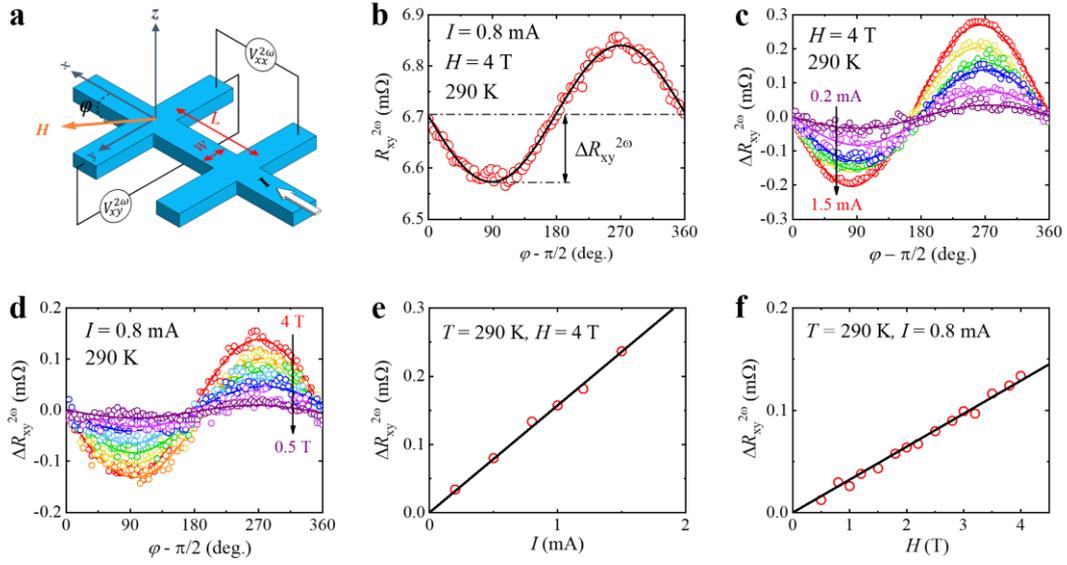

**Fig. 2 Nonlinear Hall effect measurement configuration and results. a** A schematic of the patterned $SrIrO_3$(110) sample geometry. The input current $I$ flows along the [001] direction. $\varphi$ denotes the angle between the current $I$ and the in-plane magnetic field $H$. **b** Second harmonic planar Hall resistance $R_{xy}^{2\omega} = V_{xy}^{2\omega}/I$ as a function of $\varphi$. The amplitude $\Delta R_{xy}^{2\omega}$ was extracted by fitting the $R_{xy}^{2\omega}$ data with $\cos\varphi$. **c-d** $\Delta R_{xy}^{2\omega}$ as a function of $\varphi$ under a fixed $H$ of 4 T with various $I$ from 0.2 to 1.5 mA, and under a fixed $I$ of 0.8 mA with various $H$ from 0.5 to 4 T, respectively. **e-f** The dependence of $\Delta R_{xy}^{2\omega}$ on $I$ for $H$ = 4 T, and on $H$ for $I$ = 0.8 mA, respectively.

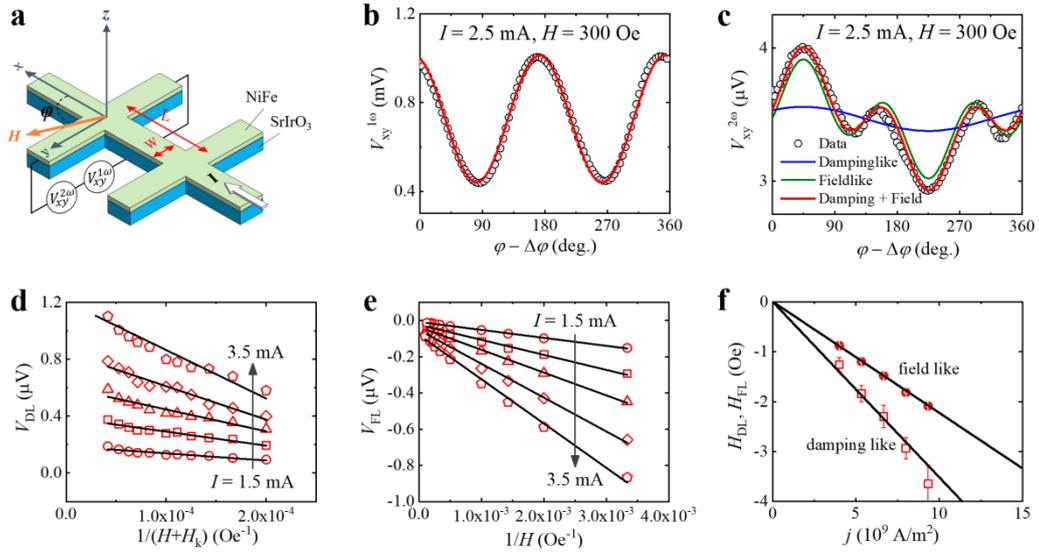

**Fig. 3 Harmonic Hall measurements of spin-orbit torques in the SrIrO$_3$/NiFe bilayer sample. a** A schematic of the patterned bilayer geometry. The input current $I$ flows along the [001] direction. $\varphi$ denotes the angle between the current $I$ and the in-plane magnetic field $H$. **b** The typical first harmonic Hall voltage $V_{xy}^{1\omega}$ as a function of $\varphi$ with $H$ = 300 Oe. **c** The typical $\varphi$ dependent second harmonic Hall voltage $V_{xy}^{2\omega}$ and the fitted curves of the dampinglike and fieldlike components. **d** The dependence of extracted dampinglike voltage $V_{DL}$ for $I$ from 1.5 to 3.5 mA on $1/(H + H_k)$, where $H_k$ is the anisotropy field of the NiFe layer. **e** The extracted fieldlike voltage $V_{FL}$ against $1/H$ for for $I$ from 1.5 to 3.5 mA. **f** The effective dampinglike and fieldlike fields, $H_{DL}$ and $H_{FL}$, as a function of the SrIrO$_3$ layer current density $J$.

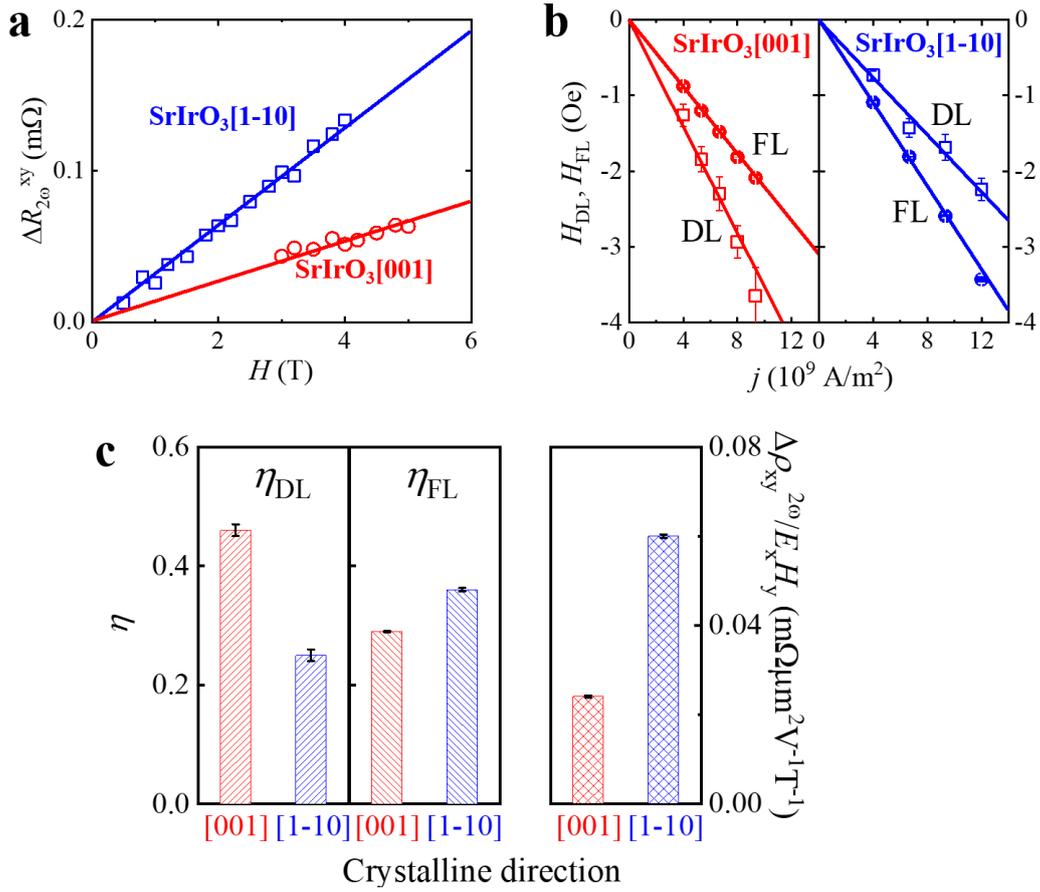

**Fig. 4 The anisotropic NHE and SOTs for the SrIrO$_3$(110). a** The NHE amplitude $\Delta R_{xy}^{2\omega}$ for the SrIrO$_3$[001] (red opened circles) and SrIrO$_3$[1-10] (blue opened squares) directions. **b** The SOT components $H_{DL}$, $H_{FL}$ as a function of the SrIrO$_3$ layer current density $J$ for the SrIrO$_3$[001] and SrIrO$_3$[1-10] directions, respectively. **c** The summarized dampinglike SOT-efficiency $\eta_{DL}$ (left column), fieldlike SOT-efficiency $\eta_{FL}$ (middle column), and the coefficient of NHE $\Delta\rho_{xy}^{2\omega}/(E_xH)$ (right column) for the [001] and [1-10] directions at 290 K.

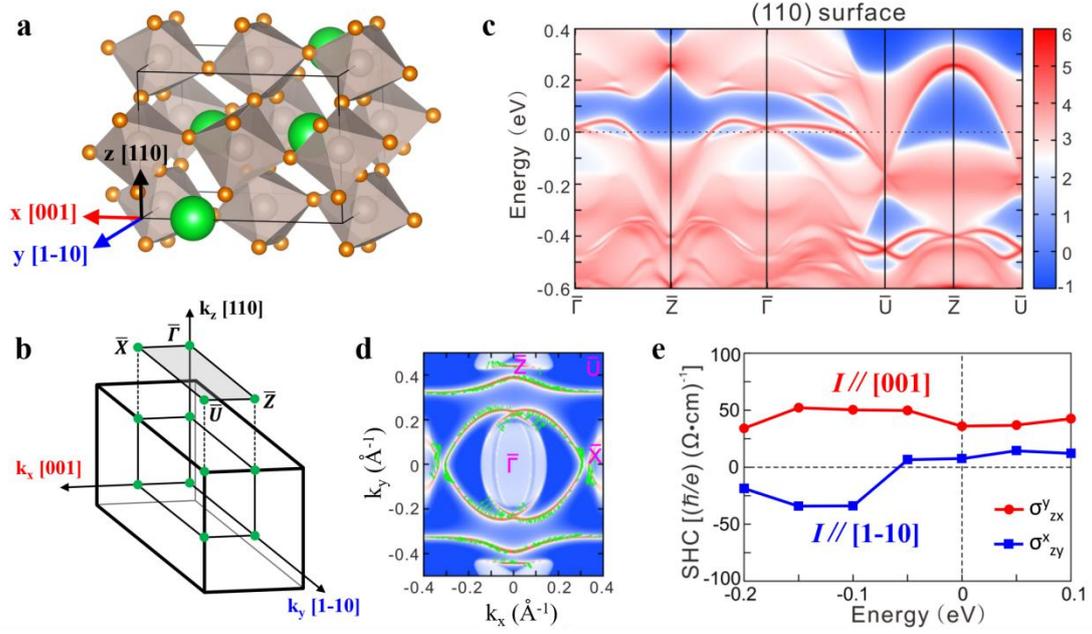

**Fig. 5 First-principles calculation results of surface states and spin Hall conductivities. a** Crystal structure of SrIrO$_3$. **b** Brillouin zone with special k points on the $k_x$-$k_y$ plane. **c** Band structure of the (110) surface. **d** Fermi surface on the $k_x$-$k_y$ plane, where the green arrows denote the spin directions. **e** Spin Hall conductivities for current *I* along the [001] (red circles) and [1-10] (blue squares) directions, respectively. Here, the notation $\sigma_{\alpha\beta}^{\gamma}$ denotes that charge current, spin current, and spin polarization are along $\beta$, $\alpha$, and $\gamma$ directions, respectively.